\documentclass[aps,epsfig,preprint]{revtex4}
\usepackage{color}
\usepackage{graphics}
\usepackage{epsfig}
\newcommand{\ts}{\textstyle}
\newcommand{\Pl}{\partial}
\newcommand{\gsim}{\raisebox{-0.6ex}{\mbox{ $\stackrel{\ts >}{\ts \sim}$ }}}

\newcommand{\bee}{\begin{equation}}
\newcommand{\ene}{\end{equation}}
\newcommand{\beea}{\begin{eqnarray}}
\newcommand{\enea}{\end{eqnarray}}

\newcommand{\fpar}[2]{\frac{{\ts \Pl \/ #1}}{{\ts \Pl \/ #2}}}

\begin{document}
\title{Propagation of Electron Magnetohydrodynamic structures in a 2-D inhomogeneous plasma}
\author{Sharad Kumar Yadav}
\email{sharad@ipr.res.in}
\author{Amita Das} 
\email{amita@ipr.res.in}
\author{Predhiman Kaw} 
\affiliation{Institute for Plasma Research, Bhat , Gandhinagar - 382428, India }
\date{\today}
\begin{abstract} 
The fully three dimensional 
governing equations in the electron magnetohydrodynamic (EMHD) regime for a plasma 
with  inhomogeneous density 
is obtained. These equations in the  two dimensional ( 2-D) limit can be cast in terms of the evolution of 
two coupled scalar fields.  The nonlinear simulations for the two dimensional case 
are carried out to understand the propagation   of EMHD 
magnetic structures in the presence of   inhomogeneity. A novel effect 
related to trapping of dipolar  magnetic structures in the high density plasma region in the EMHD regime 
is observed. The interpretation of this phenomena as well as  its relevance to the problem of hot 
spot generation in the context of fast ignition  is presented. 

\end{abstract}
\maketitle 
\section{Introduction}
The studies related to   plasma response  at fast electron time scales are becoming a topic 
of considerable research interest lately. The topics of laser plasma interaction \cite{kruer}, 
reconnection in electron current layers \cite{drake_grl}, fast Z pinches \cite{ryotov_zpinch}, 
fast plasma based switching devices \cite{pos,kingsep} and also 
fast ignition physics \cite{fi,fi_revs,kodama} primarily involve a 
study of plasma phenomena occuring at 
 electron response regime \cite{kingsep}. One particular simplified fluid model description characterizing 
the electron dynamics against the background stationary charge neutralizing ions is the 
Electron Magnetohydrodynamics (EMHD) \cite{kingsep,das_emhd}. 
The model has so far been  primarily studied  in the context 
of  plasma with uniform homogeneous density. In realistic situations, however,  the plasma  
typically has  an inhomogeneous density. The role of density inhomogeneity was 
outlined  briefly earlier in a   paper by   Kingsep \cite{kingsep}. 
The paper, however, discussed the effect in the absence of any electron inertia related 
terms. In the present manuscript the EMHD model has been generalized 
for the case of a plasma with inhomogeneous density profile including effects 
associated with finite electron inertia.  The governing equations for the generalized 
EMHD are shown to reduce to  two coupled scalar field evolution in the 2-D limit. 
For homogeneous density these equations reduce to the well known form studied 
extensively earlier \cite{das_emhd,biskamp}. The EMHD equations for the homogeneous plasma permits a 
variety of coherent localized solutions which are of a 
stationary monopolar form or a travelling (with constant velocity) dipolar form \cite{isichenko,amita_ppcf}.  
These solutions are extremely robust and have been observed to form spontaneuosly in  simulations \cite{amita08}.
Here, we study the propagation and evolution  of such coherent solutions when they 
encounter an  inhomogeneous plasma. 
This is done by  employing  our generalized EMHD  (G-EMHD) equations for the numerical evolution  
of these solutions specified as an initial condition. 


In the next section we outline  the details of 
 the derivation of the governing equations for the   
G-EMHD model corresponding to the spatially inhomogeneous plasma density. 
The 2-D limit of the equation  is then obtained and it is shown that 
the governing equations are a coupled evolution equation for two scalar fields.  
These scalar fields represent the component of the magnetic field and the vector potential 
along the symmetry direction. In 
section III we briefly recapitulate   the coherent solutions of the EMHD equations 
and their well known propagation characteristics in the context of 
a homogeneous plasma. The physical similarity of the dipole structure to a 
forward moving current pulse accompanied by spatially separated reverse shielding currents 
is also shown in this section. 
In section IV we then show with the help of our  
numerical simulations how  the density inhomogeneity influences the 
propagation of these structures. 
 The restrictions on total grid points constrains us to consider a ratio of maximum 
to minimum plasma density in our simulation domain to be at most a  factor of $10$ only. 
This constraint arises from the need of  adequately resolving  the 
 plasma skin depth scales  in the high density regime. In Section V we present 
a novel observation illustrating the trapping of dipolar EMHD structures in the 
high density plasma region. The consequence of this novel observation on the 
physics of fast ignition is also pointed out. Finally, in 
 Section VI we summarize the salient points of the paper. 
 
\section{Governing equations}
In this section we obtain the governing equations for describing 
dynamical phenomena occuring at fast electron 
time scales in the presence of plasma density inhomogeneity. The time 
scales are fast such that the ion response can be ignored. However, it is considered to be    
 slower compared to the minimum value of the local  plasma 
frequency $\omega_{pe}$  and/or $\omega_{pe}^2/\omega_{ce}$ 
(here $\omega_{ce}$ is the electron gyrofrequency) whichever is smaller. This ensures that the 
displacement current can be ignored and hence  electron density fluctuations are considered to be 
negligible.  
 This essentially implies that 
 the  electron continuity equation drops out from the evolution. Thus the system of equations which we 
derive here can be looked upon as the generalized Electron Magnetohydrodynamics (G-EMHD) model 
for inhomogeneous plasma.   

The curl of electron momentum equation can be written as 
\begin{equation}
\fpar{\vec{G}}{t} = \nabla \times \left[\vec{v}_e \times \vec{G} \right]
\label{curl_emom}
\end{equation}
where $\vec{G} = \nabla \times (m_e \vec{v}_e - {e \vec{A}}/{c})$. The electron velocity 
$\vec{v}_e = -(c/4 \pi e n_e ) \nabla \times \vec{B}$ from the Ampere's law, as the 
displacement current can 
be ignored and ions being stationary the current is determined by the electron flow 
velocity alone.  Also, here   $\vec{A}$ and $\vec{B}$ and $n_e$ 
denote magnetic vector potential, magnetic field vector and the plasma density respectively.  
We choose to normalize  the electron density with the minimum value of the plasma density 
$n_{00}$ in the region of interest, and the length by the corresponding skin depth $d_{e0} = c/\omega_{pe0}$ 
( $\omega_{pe0} = 4 \pi n_{00}e^2/m_e$ ) corresponding to this density. The magnetic field 
is normalized by $B_0$, the typical magnitude  of the  magnetic field and the time by the corresponding 
electron gyrofrequency $\omega_{ce0} = eB_0/m_ec$. The normalized equation can then be written as 
\begin{equation}
\fpar{\vec{g}}{t} = \nabla \times \left[\vec{v} \times \vec{g} \right]
\label{ncurl_emom1}
\end{equation}
\begin{equation}
\vec{v} = -\frac{1}{n} \nabla \times \vec{B}; \hspace{0.3in}
 \vec{g} = \frac{\nabla^2 \vec{B}}{n} - \nabla \left(\frac{1}{n} \right) \times (\nabla \times \vec{B})
- \vec{B} 
\label{ncurl_emom2}
\end{equation}
Equation(\ref{ncurl_emom1},\ref{ncurl_emom2})   represents the G-EMHD model written in 
normalized fields and variables. 

For the case when  the variations of the fields are confined in two dimensional $x-y$ plane, 
the G-EMHD model 
(Eq.(\ref{ncurl_emom1},\ref{ncurl_emom2}) can be  represented by a coupled set of evolution equations  
for  two scalar fields $b$ and $\psi$. These two scalar fields define the magnetic field 
of the system  through the following  relationship: 
\begin{equation}
\vec{B} = b \hat{z} + \hat{z} \times \nabla \psi
\label{bfld}
\end{equation} 
Thus $b$ represents the magnetic field component along the symmetry direction $\hat{z}$ and 
the magnetic field along $x$ and $y$ directions are $-\partial \psi/\partial y$ and 
$ \partial \psi/\partial x $ respectively. The normalized electron velocity, the expression for 
$g$, etc., in terms of these two scalar fields can then be written as   
\begin{eqnarray}
& & \vec{v} = \frac{\hat{z} \times \nabla b}{n} - \frac{\nabla^2 \psi}{n} \hat{z} \nonumber \\
& & \vec{g} = \frac{1}{n} \left[\nabla^2 b \hat{z} + \nabla^2 (\hat{z} \times \nabla \psi) \right] 
+ \nabla \left(\frac{1}{n} \right) \times 
\left[\hat{z} \times \nabla b - \nabla^2 \psi \hat{z} \right] - 
\left[b \hat{z} + \hat{z} \times \nabla \psi \right]
\label{vgfld}
\end{eqnarray} 
 Equations (\ref{ncurl_emom1},\ref{ncurl_emom2}) (which describe  the evolution of 
$\vec{g}$) may be  separated in terms of  components $g_z$ and  
$\vec{g}_{\perp}$ from which we may derive  the following coupled set of 
evolution equations for $b$ and $\psi$ fields. 
\begin{eqnarray}
\fpar{}{t}\left\{b - \nabla \cdot\left(\frac{\nabla b}{n} \right) \right\} + 
\hat{z} \times \nabla b \cdot \nabla 
\left[\frac{1}{n}\left\{b - \nabla \cdot\left(\frac{\nabla b}{n} \right) \right\}\right] 
+ \hat{z} \times \nabla \psi \cdot \nabla \left( \frac{\nabla^2 \psi}{n}\right) = 0 
\label{2db}
\end{eqnarray}
\begin{eqnarray}
\fpar{}{t}\left\{\psi - \frac{\nabla^2 \psi}{n} \right\} + \frac{\hat{z} \times \nabla b }{n} 
\cdot \nabla \left\{\psi - \frac{\nabla^2 \psi}{n} \right\} = 0 
\label{2dpsi}
\end{eqnarray}
Equations(\ref{2db},\ref{2dpsi}) represent the Generalized EMHD equations in two dimensions. 
When the  plasma density $n$ is a constant,  the above coupled set of  equations reduce to the 
 EMHD equations in 2-D.  
 
The sum of magnetic and the kinetic energy of the system is given by the expression
$$
E = \int \int \left[b^2 + \frac{(\nabla b)^2}{n} + 
(\nabla \psi)^2  + \frac{(\nabla^2 \psi)^2}{n} \right] dx dy
$$
and is an invariant for  the above coupled set of G-EMHD equations. However, when the in plane 
component of the magnetic field is zero, i.e. when   $\psi = 0$,   an additional 
square invariant  
$$
Q = \int \int \frac{1}{n}\left[b - \nabla \cdot \frac{\nabla b}{n} \right]^2  dx dy
$$
 also exists for the system. Note that for the homogeneous case  $Q-E$ represents the 
enstrophy (space integral of mean square vorticity), which is the second   invariant for the system.

In the subsequent sections we will discuss the results of the numerical simulation of 
Eqs.(\ref{2db},\ref{2dpsi}) for several distinct choices  of plasma inhomogeneity. Equations (\ref{2db}) 
and (\ref{2dpsi}) are evolved for the generalized vorticities  
$\Omega_b = L_b b = (b - \nabla \cdot (\nabla b/n))$ 
and $ \Omega_{\psi} = L_{\psi} \psi = (\psi - \nabla^2 \psi/n)$ 
using the flux corrected algorithm developed by Boris {\it et. al} \cite{boris}.  
(Here $L_b$ and $L_{\psi}$ are used to denote the operators, which upon 
operating on $b$ and $\psi$ produce $\Omega_b$ and $\Omega_{\psi}$ 
respectively.)   
The value of the $b$ and $\psi$ at each time step are then extracted from the  
generalized vorticities $\Omega_b$ and $\Omega_{\psi}$ respectively by constructing  
 matrices corresponding to  the inverse  operators $L_b^{-1}$ and $L_{\psi}^{-1}$ 
for the  chosen  spatial inhomogeneity of the  
plasma density. It should be noticed here  that the dimension of this matrix 
being extremely large ($N \times N$, where $N = Nx \times Ny$, where $Nx $ and $Ny$ represent 
the number of grid points along $x$ and $y$ directions respectively)  
it  puts a  severe constraint on the  resolution. To adequately resolve the skin 
depth at the maximum plasma density and to simultaneously also have the box dimension longer 
than  several electron skin depths (corresponding to  the minimum value of plasma density)
we were constrained to choose 
the ratio of maximum  to minimum plasma density to be at most a factor of  $10$ only.  
Alternatively, to achieve higher resolutions  a relaxation algorithm 
to evaluate $L_b^{-1} \Omega_b$ and $L_{\psi}^{-1} \Omega_{\psi}$,  
needs to be implemented \cite{storey} which utilizes the standard 
solvers for Helmholtz equation where the RAM requirements 
would not be so  extensive. We are in the process of implementing this 
for our future studies on G-EMHD model. 

\section{Coherent EMHD solutions for homogeneous plasma}
The G-EMHD equations (Eqs.(\ref{2db},\ref{2dpsi})) reduce to the following form 
for a homogeneous density $n = n_{0}$. 
\begin{eqnarray}
\fpar{\Omega_b}{t} + [b,\Omega_b] = [\psi, \Omega_{\psi}] 
\label{2db_h}
\end{eqnarray}
\begin{eqnarray}
\fpar{\Omega_{\psi}}{t} + [b,\Omega_{\psi}] = 0 
\label{2dpsi_h}
\end{eqnarray}
Here, the uniform plasma density has been chosen for the density normalization.  The 
symbol $[A,B]$ used in the equations represents a Poisson bracket between the field $A$ and $B$. 
It is then clearly evident from the above equations that 
for  the homogeneous system 
all radially symmetric (monopolar) structures are  exact stationary solutions of the equations. 
Furthermore, it  has been observed that monopolar  structures localized within a 
  spatial extent of electron skin depth are fairly robust and stable and they 
 are  spontaneously created during simulations with arbitrary initial configurations. 
A collection of monopoles which are  spatially separated by a  distance larger than the 
electron skin depth also constitute a  stationary configuration of the above 
set of equation. This is so as in  this case the self term of the Poisson bracket vanishes on account of 
radial symmetry and the cross term between two structures vanishes as there is no  spatial overlap 
amidst them. It is then interesting to see how these structures behave in the presence of density 
inhomogeneity. This is explored by us by numerically simulating the evolution of these structures 
in a given inhomogeneous density profile. The results of such studies are presented in the next section. 

The other interesting class of solutions permitted by Eqs.(\ref{2db_h},\ref{2dpsi_h}) 
have  dipolar form \cite{isichenko}. 
These solutions are not stationary but have  a steady translational velocity 
with respect to the static ion frame of the EMHD model. The dipolar solutions can be obtained 
by seeking stationarity in a moving frame $\xi = y - ut$ (assuming the translational 
velocity to be along the $y$ coordinate). Equation(\ref{2dpsi_h}) can then be written as 
the  vanishing of the Poisson bracket $ [\Omega_{\psi}, b - ux] = 0$. This implies that 
$\Omega_{\psi} = f_{\psi}(b-ux)$, where $f_{\psi}$ is a function of $b-ux$. Using this 
expression for $\Omega_{\psi}$ as well as the stationarity criteria in the $\xi$ frame,  Eq.(\ref{2db_h}) 
can  be expressed as  
$$
[\Omega_b + f_{\psi}^{\prime} \psi, b-ux] = 0; 
$$
Here $^{\prime}$ indicates a differential with respect to the argument $(b - ux)$ of the function $f_{\psi}$. 
This suggests that $\Omega_{b} + f_{\psi}^{\prime} \psi = f_b(b-ux)$ 
($f_b$ being another function of $b-ux$). Thus a travelling solution can be obtained by seeking 
solutions of the following coupled set: 
\begin{eqnarray}
& &\nabla^2 \psi - \psi = f_{\psi}(b-ux) \nonumber \\
& &\nabla^2 b - b + f_{\psi}^{\prime} \psi = f_b(b-ux) 
\label{dipole}
\end{eqnarray} 
The  general solutions  would correspond to  any choice of the functions $f_b$ and $f_{\psi}$. 
However,  analytical form of the dipole solutions are  typically obtained by choosing the vorticity 
source functions $f_b$ and $f_{\psi}$ 
as linear functions of their argument $(b-ux)$ in the  inner  spatial region of 
 radii $r \le r_0$ around the centre of the structure. The differential equation can be separated in 
$r$ and $\theta$ coordinates in the two dimensional $x- y$ plane. It can be shown that 
an explicit appearance of $x \sim r \cos(\theta)$ (as a coefficient of $u$)  produces a $\cos(\theta)$ 
dependence and hence a dipolar form of the solution. Furthermore for the radial part in this inner region,  
 the solutions can be represented in terms of a  Bessel function of the  first kind denoted 
by $J$. To achieve localization 
the vorticity source functions $f_b$ and $f_{\psi}$ are 
 chosen to be zero in the outer spatial region  $r > r_0$. The solutions for the fields 
$b$ and $\psi$ in the outer  region are thus  modified Bessel function $K$ of the radial coordinate.   
 The matching  of the field and its 
derivative in the inner and outer region yields a localized  solution.  

It might  appear that the specific linear choice made 
for the inner region is too restrictive and may in general not correspond to reality. 
However,  it has been shown recently  (for the system with $\psi = 0$) 
that a linear choice of $f_b$ is 
consistent with a self organization paradigm based on the enstrophy minimization subject to the 
constancy of the energy of the system \cite{amita08}. The  approach of the system to 
such a self organized state has been confirmed by  
 numerical simulations which show 
 the spontaneous  formations of  structures from an arbitrary initial condition 
satisfying Eq.(\ref{dipole}) for a linear functional form of $f_b$ \cite{amita08}. 
Similar  studies in the presence of  finite   $\psi$ 
field are underway and the results will be reported elsewhere. 
In the present manuscript we report the propagation characteristics  of the dipolar solutions for both cases 
(viz. $\psi = 0$ and $\psi$ finite, with  $f_b$ and $f_{\psi}$ both as linear functions of the argument $b-ux$) 
as they encounter the inhomogeneous plasma density profile. 

It should be noted that a  monopolar structure is like a circularly rotating  current similar to that of a  
solenoid. The propagating dipoles on the other hand can be viewed as an  
electron current pulse translating in space.    
The central spatial region within the 
positive and the negative peaks of the magnetic structure carries electron current in the direction 
of the propagation of the dipole, whereas  the outer edge region carries a reverse current as 
shown in the schematic plot of Fig.1. Thus the dipole structure can be viewed as a current pulse 
containing a spatially separated forward and the reverse shielding electron current pulse. 
 The current configuration of a 
dipole is thus quite similar to the one encountered during the ignition phase of the 
fast ignition experiments. The forward energetic electrons generated at the critical density 
surface moves towards the dense target core. The background plasma in this case 
provides for the return shielding current. The two currents get spatially separated 
through Weibel instability and form current channels in which the inner region carries 
the forward current and the outer region the return shielding current. 
The numerical study of the propagation of the dipole magnetic structure through an inhomogeneous 
plasma and its behaviour in the high density region is therefore an important  issue for investigation. 
In the subsequent sections the result of such a simulation will be presented.

\section{Numerical studies} 
In this section we study the propagation of both monopolar and dipolar solutions through an 
inhomogeneous plasma. The plasma density inhomogeneity is chosen to have the following 
spatial profile 
\begin{equation}
n(x,y) = a - b \tanh \left\{\frac{\sqrt{y^2} - w}{\sigma} \right\}
\label{profileA}
\end{equation}
The parameters $a$ and $b$ were chosen appropriately 
to define either a density hump in the region $\mid y \mid \le w $ or a density cavity. 
For the above chosen density profile the constant density contours form straight lines parallel 
to the $x$ axis. 


\subsection{Propagation of monopoles}
The monopolar structures are the exact stationary equilibrium solutions of the homogeneous 
EMHD equations. In order to study the influence of plasma inhomogeneity on their evolution 
we place a  monopolar structure initially at the boundary region (around $y \sim {w}$), 
where the plasma density inhomogeneity is typically high. The subplots of the Fig.2 
show the evolution of this structure. For this case we had chosen  
the simulation box  of size $L_x = L_y = 10$ and 
$x$ and $y$ coordinates  range from $-5.0$ to $5.0$. For the plasma density we 
had chosen  $a = 5.5$, $b = 4.5$, $w = 2.5$ and $\sigma = 1.0$.  
The maximum and minimum value of density is therefore $n_{max} = 10$ and 
$n_{min} = 1$ respectively. The local electron skin depth therefore ranges from $0.3 \le d_e \le 1.0$. 
 The high density plasma region here is confined within   $ \mid y \mid \le w$ for all $x$. 
The density falls sharply within a length $\delta y = {\sigma}$  
from $10$ to unity beyond $\mid y \mid \gsim {w}$. The density scale 
length at the boundary is typically of the order of the electron skin depth as 
$\sigma = 1.0$.

From the figure it is clear that the monopolar 
structure moves transverse to the density gradient. The monopoles being stationary 
in the homogeneous plasma the propagation velocity is clearly an artifact of the 
presence of plasma density inhomogeneity. 
The direction as well as the magnitude of the propagation velocity is observed to match with the  
following  drift velocity. 
\begin{equation} 
\vec{V}_n = \frac{b \hat{z} \times \nabla n}{n^2} 
\label{vdrift}
\end{equation}
which can be obtained from the Eq.(\ref{2db}) upon ignoring electron inertia related terms. 
For  the density profile of Eq.(\ref{profileA}) $n$ is a function of $y$ alone and we have 
the propagation  along $x$ and the magnitude of the velocity is 
$$
V_{nx} = b \fpar{}{y}\left(\frac{1}{n} \right)
$$ 
From the subplots of the Fig.2, the value of $V_{nx}$ evaluated by observing the 
distance propagated by the structure along $x$ is  $0.0307$ which is 
close to that  estimated from the above  expression for the 
electron drift velocity, as $b$ typically ranges from $ 0.0233$ to $0.1997$ in the monopolar 
structure and $\partial (1/n)/\partial y $ ranges from $0.1131$ to $0.448$  over the structure. 
This implies that the value of 
 $V_{nx}$ from the expression can be about $ 0.0026$ to $0.089$. The observed value lies within this range. 
In fact the average of $V_{nx}$ evaluated over the $y$ extent of the  structure (through which the structure 
would translate) turns out to be 
very close  $0.0369$   to the observed velocity. This clearly indicates that the monopole is essentially 
propagating with the drift velocity of $<V_{nx}>$. 

The  other density gradient dependent terms arising through the  finite electron inertia 
related terms are typically  smaller in magnitude and they generally contribute  as a  
source causing modification of  the spatial profile of the  magnetic structure. 

\subsection{Propagation of dipoles} 
We next study the behaviour of dipoles as they encounter the region of plasma 
density inhomogeneity. Since the dipolar structures are known to propagate 
along their axis, we start our simulation by placing   
 an exact nonlinear dipole  structure at some distance away from the density gradient region, i.e. 
at a location of a uniform low density plasma where $n = 1$  at  a positive $  y >  {w}$. The sign 
of the lobes in this case are chosen in such a way that the dipole propagates towards the 
high  density region. We observe  that the dipolar structure 
crosses past the inhomogeneous  density region to enter the high plasma density region.    
The subplots of Fig.3 clearly illustrate the penetration of the dipolar structure in the high 
density plasma region. (For the plots of this figure we again have 
$a = 5.5 $, $b = 4.5$, $w = 2.0$ and $\sigma = 0.4$. The box length  in this case is 
$L_x = L_y = 4\pi$.)  
In fact we observe that  at the inhomogeneous density 
region the axial translational velocity of the dipole increases considerably.  The two lobes  
get squeezed towards each other forming a shock like structure 
in the direction transverse to the density gradient. 
This behaviour appears to be in stark contrast to  the propagation characteristics of 
the  monopolar structures,  which merely show a movement transverse to the density gradient 
direction.  

The observations on dipole propagation   can, however,  be understood readily. For the dipole structure 
approaching the high density plasma region 
(along decreasing $y$, in  Fig.3) the  
 left lobe corresponds to  positive values  of $b$ whereas the  right lobe has negative $b$ values. 
Clearly, when the two lobes of the dipole encounter the density inhomogeneity the left 
lobe has a  drift velocity due to the density inhomogeneity towards  right (positive $x$ direction) 
whereas the right one drifts towards the negative $x$ direction. This sqeezes  the two lobes 
of the dipoles closer in $x$. As a result the  size of the lobes as well as 
their separation  gets significantly reduced.  
This also causes an  enhancement  in the magnitude of $\mid b \mid$ of the  two lobes. 
The reduced distance  between the lobes as well as the enhanced amplitude of $\mid b \mid$ 
results in an  increased  axial propagation velocity of the dipole. This accelerates the 
penetration of the dipolar structure in the high density plasma region. 
 It should be noted that for the  case where the dipole approaches a    density cavity 
the effect is entirely different. The sign of $\nabla n$ being opposite, in this case the lobes 
 separate   with the drift speed. The 
separation results in a reduced axial velocity of the dipole, which ultimately 
dimishes to zero as the separation distance between the two lobes 
exceeds the electron skin depth distance. The two lobes then separately move as monopolar 
structures transverse to the density gradient direction. Thus, the 
 dipole is   unable to penetrate the region of lower plasma density. Our simulations 
indeed show this effect as the subplots of Fig.4 clearly illustrate. Here the central region 
$\mid y \mid \le w$ corresponds to a low density plasma region with $n = 0.0.2$. 
Here $a = 0.6$, $b = -0.4$ and other parameters are same as that of Fig.3.

Let us now study in detail the behaviour of the dipole as it 
 enters the high density plasma region. 
Though the shape of the dipole is considerably distorted while it traverses the 
inhomogeneous plasma region, but once it is inside the  high density homogeneous plasma 
region it regains  the familiar dipolar form. The scale length of the dipole, in the 
high density region changes  by the same factor as the ratio  of the skin depth  
of the high and low density regions. For instance   the initial 
dipole was chosen to have $r_0 = 1.0$ and at $t = 690$ when it is completely inside the high density region 
the value of $r_0 = 0.47$ (a reduction by a factor of  approximately $1/3$). 
We thus observe that the dipolar structures are fairly robust. Even after encountering a strong 
density inhomogeneity, once in the region of homogeneous 
plasma  they adjust smoothly to the new value of the density. 

In the next section we illustrate an interesting consequence of the 
above observations, namely that of   
novel trapping behaviour of  EMHD dipoles   in a high density plasma region.

\section{Trapping of EMHD dipoles in high density plasma}
The propagation characteristics of the  dipolar structure 
discussed in the preceding section  has an interesting consequence.  It suggests 
that the EMHD magnetic structures of dipolar form can enter a high density plasma region. However, once 
inside a high density plasma region they would remain trapped there. 

We choose  a density profile $n(x,y)$ which has a high density region 
with a finite transverse extent   as well.  A  circular region in the $x-y$ plane is 
chosen to have a high density of the plasma. The functional form of the plasma density is 
\begin{equation}
n(x,y) = a - b \tanh \left\{\frac{\sqrt{x^2+y^2} - w}{\sigma} \right\}
\end{equation}
The choice of parameters for simulation with this density profile is 
 $a=5.5; b=4.5; w=2.0; \sigma=0.4$. For this particular density profile a 
 dipole is placed with its centre on 
 the line $x = 0$ at the positive value of $y = 4.0$.  The  axis of the dipole 
is parallel to  the $y$ axis as can be seen from the  
subplot at $t = 0$ of Fig.5. The dipole velocity is directed along the negative 
$y$ axis so that it approaches 
 the high density plasma region. It can be seen from the subsequent subplots that due to 
the individual density inhomogeneity related drift velocity of the two lobes,  
the two lobes of the dipole approach each other.  This enhances the 
axial dipolar velocity and the   
dipole structure  enters the high density region. Once inside the 
homogeneous high density region it  
translates along its axis which is along the 
diameter of the circular high density region.  Upon reaching the other end  the  
dipole again encounters the inhomogeneous plasma density region. However,  the direction 
of the density gradient  is now opposite to  the one that it 
encountered while entering the high density region. Thus in this region 
 the two lobes of the dipole separate from each other. As the separation  between the lobes exceeds the 
skin depth distance the lobes act like individual monopolar structures and move transverse to the density 
gradient. In this case the density gradient being along the radial direction, the two structures 
 move along the perimeter of the circle. They thus again come in close contact at the topmost 
point of the circle from where  they had entered  the high density region. At this place they again form a 
dipolar structure and translate along the diameter of the high density region. 
The simulations clearly show the repetition of this  cycle. 
It is thus  clear that the dipole structure remains trapped inside the 
high density plasma region.

These numerical  results  
showing  penetration of the dipolar structures inside the high density region 
and its trapping within the high density  region suggests an important implication for the problem 
of fast ignition \cite{fi}. In this scheme one relies on energetic electrons generated 
at the critical density surface by the ignitor laser pulse to penetrate the high density 
pre-compressed target core and deposit its energy there for the creation of hot spot. 
Here, we have  shown that a current pulse in the form of a dipolar structure 
(similar in form to  a forward moving  electron current  and a spatially separated reverse shielding current) 
 can  easily penetrate a high density plasma region. It has also been observed that  once 
inside the compressed target  the structure is unable to  come  out of  the compressed 
high density region. This suggests  that  there will 
be ample time for them to dump their energy in the high density 
compressed region of the core of the fusion target plasma.



Another interesting thing to note is the fact that as the dipole enters the high density 
region the squeezing of its lobes results in a shock  formation along the direction transverse 
to the density gradient. Such a shock formation 
is an artifact of the thermoelectric like source term $\nabla P \times \nabla n$ in the  magnetic 
field evolution equation where the self consistent magnetic field $B^2$, itself acts as a   
pressure term. This nonlinearity is responsible for the shock formation. The sheared current 
layer in the magnetic shock structure would lead to the Kelvin Helmholtz (KH) like instability \cite{adas}. 
The instability converts the 
electron flow energy into fine scale vortices. In the three dimensional case such 
vortex flows   cascade energy
towards finer scales \cite{njain} and eventually dissipate into heat by electron Landau damping effects.  
We  would show through simulations as well as analytical modelling in a 
subsequent publication that the shoch width adjusts in such a fashion that 
the energy dissipation in such a shock structure is essentially 
independent of the value of the dissipation mechanism. This again is very important 
in the context of  fast ignition scheme as it suggests that the energy dissipation would occur 
no matter how small is the value of the classical collisional damping process in the dense target core.

Furthermore, in some recent fast ignition experiments it has  been observed 
that if a  metallic wire is attached to the 
gold cone where the ignitor laser pulse is incident, the energetic electrons generated at the 
critical density surface choose a 
 preferential  path provided by the ionized wire to traverse from the low density plasma corona towards the 
compressed target core \cite{kodama_wire}. 
This result is important as it indicates that the energetic electrons can be guided 
in the low density plasma corona region to reach the dense target core. An understanding 
of this pheneomena has, however, been lacking. We feel that this can again be understood on the 
basis of the novel trapping mechanism outlined above. 
The ionized wire in this case forms  a high density plasma path amidst the 
 low density coronal plasma.  The high density region can then  trap  the current pulse structure formed 
by the combination of 
the forward energetic electrons and the background reverse shielding current.

\section{Summary }
A generalized Electron Magnetohydrodynamic (G-EMHD)  model has been introduced here which accounts 
for the fluid model description of fast electron time scale phenomena in an inhomogeneous plasma. 
The electron density fluctuations have, however, been ignored by restricting to 
times scales which are slower than electron  plasma period $\omega < \omega_{pe}$ and/or 
$\omega < \omega_{pe}^2/\omega_{ce}$ whichever is smaller. This is the 
same condition which is invoked for the derivation of the EMHD model and justfies ignoring the 
displacement current. A reduced G-EMHD model for 2-D case has also been  obtained. The model is 
represented in terms 
of a coupled set of  evolution equation for  
two scalar fields. The two scalar fields correspond to the component of the 
magnetic field and the vector potential along the symmetry 
direction. 

The G-EMHD equations in 2-D were used to study numerically the evolution and propagation of 
nonlinear  coherent solutions of the EMHD equations in the presence of density inhomogeneity. 
The two varieties of coherent solutions (viz., the stationary monopolar solutions and the 
travelling dipolar solutions) show interesting novel aspect of propagation as they encounter 
a region of plasma density inhomogeneity. The monopolar structures are no longer stationary but 
are observed to propagate with a diamagnetic drift velocity. The dipoles on the other hand 
display a very novel behaviour. They 
show penetration inside the high density plasma  region from a lower density side. 
However, once inside the higher density region they get  trapped inside this region. 
It has been shown that this behaviour of trapping  
 can have interesting favourable implications in the context of the 
energetic electron propagation and their stopping inside  the dense target core 
in the context of fast ignition experiments.

\noindent
{\bf{Acknowledgement:}} This work was financially supported by DAE-BRNS sanction 
no.: 2005/21/7-BRNS/2454. SKY and AD would like to thank Sarveshwar Sharma for his help in 
the preparation of some of the figures.

\newpage
\begin{center}
FIGURE CAPTIONS
\end{center}
\vspace{0.2 in}
\begin{itemize}
\item[Fig.1]
A schematic plot showing the dipole structure (subplot(a)) represented by constant 
contour lines in the $x-y$ plane for  the magnetic field $b$  along the symmetry direction.  
The subplot(b) shows a plot of  the magnetic field amplitude  as a function of $x$ at the 
mid $y$ plane of the dipolar structure. In subplot (c) the corresponding electron current 
is shown. The central region represents the forward electron current (electron current along the propagation 
direction of the dipole) and the reverse electron current flows at the edges. 
 \item[Fig.2]
(Color Online) 
The propagation of the monopolar structure ( color contours ) in a inhomogeneous plasma density is depicted by 
showing the location of the structure  at various times in the different 
subplots of the figure. The thick black lines represent the plasma density contour. In this case the 
plasma density is chosen to be a function of $y$ only. The central $y$  region 
of width $w = \pm 2.0$ corresponds to a high density ($10$ times the density at the edge ( $\mid y \mid \ge 2.0$).  
The monopole structure is seen to propagate transverse to the density gradient.
 \item[Fig.3]  
(Color Online) Various stages of the propagation of a dipolar structure 
through an inhomogeneous desnity plasma has been shown. 
The inhomogeneity in plasma density is similar to that of Fig.2 in this case. The figure clearly
shows the penetration of the dipole through the plasma density inhomogeneity to enter the high density region. 
The lobes of teh dipole structure are squeezed towards each other as they pass thorugh the inhomogeneous 
region. However, once inside the high density homogeneous region they again aquire a balanced dipolar form. 
\item[Fig.4] 
(Color Online) In this figure the dipole is shown to approach a density cavity (lower density plasma region). 
It can be observed that the dipole is unable to penetrate the lower density plasma. The two lobes of the dipole 
get separated transverse to the density gradient direction and subsequently 
they evolve as separate monopolar structures.  
\item[Fig.5] 
(Color Online) The trapping of the dipolar structure in a high density plasma has been illustrated in this figure. 
A high density plasma with a circular profile in the $x - y$ plane represented by the thick black 
contour lines are depicted on the various subplots. A dipole structure can be seen to penetrate the 
high density region. However, once inside the high density region it continues to remain rapped 
in this region. 

\end{itemize}

\end{document}